\documentclass{article}

\usepackage{arxiv}            

\usepackage[utf8]{inputenc}
\usepackage[T1]{fontenc}
\usepackage{graphicx}
\usepackage{multirow}
\usepackage{amsmath,amssymb,amsfonts}
\usepackage{booktabs}
\usepackage{textcomp}
\usepackage{xcolor}
\usepackage{siunitx}
\usepackage{float}            
\usepackage{natbib}          
\usepackage{url}
\usepackage{hyperref}
\hypersetup{colorlinks=true, linkcolor=blue, citecolor=blue, urlcolor=blue}

\title{Frictional timescales and the impact of climate change-driven extreme weather on rainfall-triggered landslides in Mizoram, NE India}

\renewcommand{\shorttitle}{Frictional timescales of Mizoram landslides under climate extremes}

\author{%
  Pritom Sarma\textsuperscript{1,2}\thanks{Corresponding author: \texttt{pritom.sarma@mail.huji.ac.il}} \quad
  Krishnendu Paul\textsuperscript{3} \\[0.6em]
  \textsuperscript{1}\,Institute of Earth Sciences, Hebrew University of Jerusalem, Jerusalem, Israel \\
  \textsuperscript{2}\,Department of Earth and Environmental Sciences, Tulane University, New Orleans, Louisiana, USA \\
  \textsuperscript{3}\,SRK Consulting Asia Limited, Kolkata, India
}

\date{}

\begin{document}
\raggedbottom
\maketitle

\begin{abstract}
Mizoram records the highest landslide frequency among all Indian states, yet physics-based models that predict the \emph{timing} of slope failure remain unavailable for the region. Here, we apply the rate-and-state friction (RSF) block-slider framework of Paul et al.\ (2024) to 19 rainfall-triggered landslides in and around Aizawl (2016--2025) to investigate the hydro-mechanical coupling between pore-pressure transients and the slow-to-fast transition of slopes hosted in Miocene shale-dominated lithology. For each event, satellite-derived (GPM) rainfall is propagated to failure depth using a 1D infiltration model across three hydraulic conductivity scenarios, and RSF parameters are inverted to reproduce the observed failure date. The resulting dimensionless normalized pore-pressure $\chi = \mu_0 \Delta P / (a\,\sigma'_f)$ cleanly separates the 19 events into two dynamically distinct failure regimes: synchronous failure ($\chi \gtrsim 4$, zero delay from peak pore-pressure), exemplified by the eight-event Cyclone Remal cluster of May 2024, and delayed failure ($\chi \sim 1$--$3$, delays of hours to 10~days), controlled jointly by $\chi$ and the velocity-weakening ratio $a/b$. Using CMIP6 extreme-rainfall scaling factors for northeast India under SSP2-4.5, SSP3-7.0 and SSP5-8.5 scenarios, we project that the fraction of landslides falling in the zero-warning synchronous regime increases from the current $\sim$56\% to $\sim$72\% under SSP5-8.5. Our results imply a significant climate-driven escalation of multi-site, clustered landslide failure risk driven by the intensification of extreme precipitation events.
\end{abstract}


\section{Introduction}\label{sec:intro}

Landslides constitute one of the most pervasive geohazards across the hilly terrains of India's northeastern region, where a combination of fragile geology, intense monsoon rainfall, and accelerating anthropogenic modification of slopes produces thousands of slope failures annually \citep{Martha2023,Froude2018}. Among the eight northeastern states, Mizoram stands out: the National Remote Sensing Centre (NRSC) Landslide Atlas of India documented 12{,}385 landslide events in Mizoram between 1998 and 2022, making it the most landslide-prone state in the country \citep{Martha2023}. The capital city Aizawl, home to over 400{,}000 residents and built on an anticlinal ridge of Tertiary interbedded sandstone and shale \citep{Lallianthanga2013}, is acutely vulnerable: over 44\% of the city area has been classified as unconditionally unstable by hillslope stability analysis, and more than 300 landslide-related fatalities have been recorded in the past decade alone \citep{SangiEtAl2025}.

The devastating impact of Cyclone Remal in May 2024, which triggered at least eight simultaneous landslides in Aizawl killing 34 people in a single day, brought new urgency to the problem \citep{SangiEtAl2025}. This event exemplified a comparatively rare, high-intensity mode of failure relative to the cumulative monsoon-driven landslides that have historically characterized the region: a single extreme rainfall pulse of approximately 205~mm in 24~hours overwhelmed many slopes near-simultaneously, leaving little to no warning interval between the onset of heavy rain and catastrophic failure. As climate projections for northeast India consistently indicate intensification of extreme precipitation events even as mean monsoon rainfall declines \citep{DharaEtAl2025,SahaEtAl2023,PalEtAl2025,Roxy2017}, understanding the mechanistic controls on whether a slope fails synchronously with peak rainfall or days later becomes a question of direct societal relevance.

While landslide susceptibility maps and empirical rainfall intensity-duration thresholds are available for several Indian regions \citep{Abraham2020,Caine1980,GuzzettiEtAl2008,SegoniEtAl2018}, no physics-based model of landslide failure \emph{timing} has been applied to Mizoram. Susceptibility maps identify \emph{where} landslides are likely and rainfall thresholds identify \emph{how much} rain is needed, but neither addresses \emph{when} failure occurs relative to the causative rainfall, a critical gap for early warning and evacuation decisions. The present study targets this \emph{when} question directly through a physics-based failure-timing framework (Section~\ref{sec:methods}).

Conventional limit-equilibrium and Mohr--Coulomb stability analyses assess \emph{whether} a slope is stable through a static factor of safety but, being time-independent, cannot describe the rate at which a destabilizing slope accelerates towards failure \citep{Iverson2000}. Capturing failure \emph{timing} therefore requires a rate-dependent constitutive description of the basal shear zone. Rate-and-state friction (RSF) theory, originally developed to describe the velocity-dependent frictional strength of rocks in the context of earthquake mechanics \citep{Dieterich1979,Ruina1983,RiceRuina1983,Marone1998,Scholz1998}, provides such a framework, capable of predicting failure timescales. An increasing body of evidence demonstrates that landslide basal shear zones exhibit RSF-like frictional behaviour, including rate-dependent friction and memory-dependent state evolution \citep{ScaringiEtAl2018,KangEtAl2022,Helmstetter2004}. Building on this, \citet{Handwerger2016} showed that a single RSF model captures both slow sliding and catastrophic failure of landslides, and \citet{PaulEtAl2024} developed a comprehensive analytical treatment of the slow-to-fast transition timescales for RSF-governed slopes driven by pore-pressure transients. Their key finding is that the failure regime is governed by two dimensionless quantities (a normalized pore-pressure comparing the rainfall-driven pore-pressure rise to the slope's frictional resistance, together with the velocity-weakening ratio of the basal shear zone), which jointly determine whether a slope fails synchronously with peak pore-pressure or only after an extended period of pre-failure creep (both quantities are formally defined in Section~\ref{sec:rsf}). This framework was validated against 19 Indian landslides in the original study, but it has not yet been applied to understand extreme, cyclone-driven triggering or to project how failure timing may evolve under a warming climate.

Here, we apply the \citet{PaulEtAl2024} framework as a meta-analysis across 19 landslides in and around Aizawl spanning 2016--2025. Our objectives are threefold: (1)~to characterize the RSF parameter space of Aizawl's landslide-prone slopes and test whether the theoretically predicted bimodal failure regime is borne out across a diverse set of real events, (2)~to analyse the Cyclone Remal cluster as a natural experiment in which cyclone-remnant rainfall overwhelms frictional resistance, and (3)~to evaluate how projected changes in extreme rainfall intensity under Coupled Model Intercomparison Project Phase~6 (CMIP6) scenarios may systematically shift the balance from delayed to synchronous failure, with implications for early warning capability.

\section{Study area and geological setting}\label{sec:studyarea}

Aizawl (23.73$^\circ$N, 92.72$^\circ$E; elevation $\sim$1{,}188~m~a.s.l.) lies within the Tripura--Mizoram Fold Belt (TMFB), the outer accretionary wedge of the Indo-Burma subduction system where the Indian Plate subducts obliquely beneath the Burma Plate \citep{Nandy1986}. The TMFB consists of a series of N--S trending anticlines and synclines with progressively younger deformation from east to west. Aizawl itself is built on a NE--SW trending anticlinal ridge underlain by the Bhuban and Bokabil Formations of the Surma Group (Miocene age), comprising monotonous sequences of interbedded sandstone, siltstone, and shale (described as a ``great flysch facies'') with bedding dips of 20$^\circ$--70$^\circ$ \citep{Tiwari2011}. The Bhuban Formation, the thickest lithostratigraphic unit in Mizoram at approximately 5{,}000~m, is dominantly argillaceous in its middle portion \citep{Lallianthanga2013}. This interbedded sequence creates an inherently failure-prone configuration: massive sandstone beds cap hilltops with sufficient porosity and permeability for rainfall infiltration through widely spaced orthogonal joints, while water accumulates at impervious clayey shale contacts, causing considerable reduction of shear strength at these interfaces \citep{SardanaEtAl2019}. Differential weathering further destabilizes slopes as shale layers weather rapidly to a soil-like consistency while sandstone maintains structural rigidity, creating overhanging blocks and potential slip surfaces.

Published geotechnical data from the Hunthar Veng landslide investigation quantify this weakness: surficial debris has cohesion of only 20~kPa and friction angle of 11$^\circ$ at unit weight 20~kN\,m$^{-3}$, while the underlying highly weathered rock has back-analysed strength parameters $c = 100$~kPa and $\phi = 38^\circ$ \citep{VinothEtAl2020}. Under saturated conditions, slopes were found unstable even with engineered nail interventions (factor of safety only 1.02--1.05). Kinematic analysis indicates predominantly planar and wedge-type failure mechanisms \citep{SardanaEtAl2019}. The region falls within India's highest seismicity zone (Zone~V) and receives $\sim$2{,}094~mm mean annual rainfall, approximately 90\% during the southwest monsoon (June--September) \citep{Lallawmzuali2024}. Anthropogenic factors (unplanned urbanization, construction without geological assessment, slope cutting for roads, and inadequate drainage) severely compound the natural instability.

\begin{figure}[H]
\centering
\includegraphics[width=0.9\textwidth]{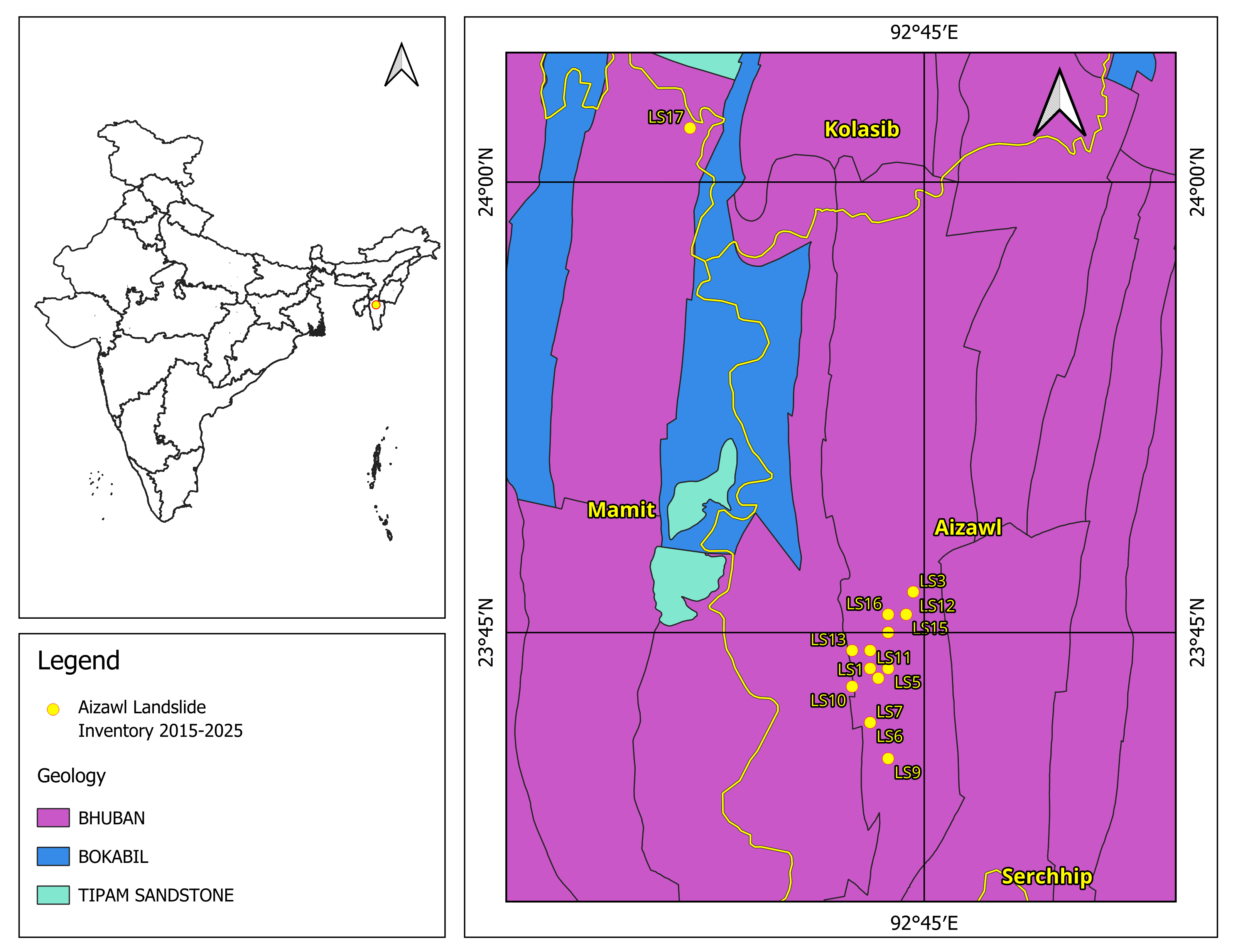}
\caption{Geological map of the Aizawl region showing the locations of all 19 inventoried landslides (yellow dots) overlaid on the Bhuban (pink), Bokabil (blue), and Tipam Sandstone (teal) formations. District boundaries (Kolasib, Mamit, Aizawl, Serchhip) and major roads (yellow lines) are shown. LS17, located near Kawnpui on the Aizawl--Kolasib border, is the only event outside the main Aizawl cluster. Inset: location of the study area within India}\label{fig:studyarea}
\end{figure}

\section{Landslide data}\label{sec:data}

We compiled an inventory of 19 rainfall-triggered landslides in and around Aizawl spanning nearly a decade (September 2016 to May 2025). The inventory draws on three categories of sources: (i)~published scientific literature \citep{VinothEtAl2020, SangiEtAl2025}; (ii)~the Geological Survey of India (GSI) Bhukosh/Bhusanket landslide incidence database (\url{https://bhukosh.gsi.gov.in}); and (iii)~contemporaneous media reports cross-verified against government situation reports from the Mizoram State Emergency Operation Centre (SEOC) and the Department of Information and Public Relations (DIPR).

Table~\ref{tab:inventory} summarizes the inventory and Fig.~\ref{fig:studyarea} shows the spatial distribution of all 19 events on a geological basemap. The events encompass a range of failure types (debris slides, translational slides on dipping shale beds, rockslides, quarry collapses, debris flows, retaining-wall failures, and road blockages), all hosted within the Bhuban Formation. All are classified as shallow planar failures with reported or inferred failure depths of $\sim$2~m. The events are triggered by two distinct meteorological forcing types: (a)~cumulative monsoon rainfall (11~events spanning 2016--2025), and (b)~extreme rainfall from the remnants of Cyclone Remal (8~events on a single day, May 28, 2024), which delivered $\sim$205~mm to Aizawl within 24~hours \citep{SangiEtAl2025}. Because a single, identical meteorological forcing acted on multiple slopes with potentially different frictional properties, this cyclone-triggered cluster provides a natural comparison for isolating the slope-intrinsic frictional controls on failure timing. The cluster accounted for 34 of the 47 fatalities (72\%) recorded across the inventory. Coordinates are reported as published or as approximate centroids based on locality descriptions; coordinate quality is noted in the supplementary material. Notably, LS4 from the inventory was excluded from the RSF analysis because it was classified as a retaining-wall failure without a natural slip surface, leaving 18 events with natural failure mechanisms (referred to as 19 including LS4 in Table~\ref{tab:inventory} for completeness of the inventory record).

\begin{table}[H]
\centering
\caption{Landslide inventory and site characteristics for the Aizawl region (2016--2025). All events are shallow planar failures at $\sim$2~m depth in interbedded sandstone--shale of the Bhuban Formation}\label{tab:inventory}
\footnotesize
\begin{tabular}{@{}llllrrl@{}}
\toprule
ID & Date & Location & Lat., Long. & Fat. & Type & Trigger \\
\midrule
LS1  & 2016-09-17 & Tlangval           & 23.73, 92.72 & 5  & Debris slide   & Monsoon \\
LS2  & 2017-09-17 & College Veng       & 23.72, 92.72 & 0  & Shale failure   & Monsoon \\
LS3  & 2019-07-02 & Durtlang Leitan    & 23.77, 92.74 & 3  & Transl.\ slide  & Monsoon \\
LS5  & 2021-06-13 & Ngaizel, Kulikawn  & 23.73, 92.73 & 0  & Rockslide       & Monsoon \\
LS6  & 2024-05-28 & Melthum quarry     & 23.70, 92.72 & 15 & Quarry collapse & Cyc.\ Remal \\
LS7  & 2024-05-28 & Hlimen             & 23.70, 92.72 & 6  & Debris slide    & Cyc.\ Remal \\
LS8  & 2024-05-28 & Salem Veng         & 23.74, 92.72 & 3  & Bldg.\ collapse & Cyc.\ Remal \\
LS9  & 2024-05-28 & Falkawn            & 23.68, 92.73 & 2  & Debris slide    & Cyc.\ Remal \\
LS10 & 2024-05-28 & Chawnpui           & 23.72, 92.71 & 8  & Debris slide    & Cyc.\ Remal \\
LS11 & 2024-05-28 & Republic Veng cem. & 23.73, 92.72 & 0  & Mult.\ slides   & Cyc.\ Remal \\
LS12 & 2024-05-28 & BSUP, Durtlang     & 23.77, 92.74 & 0  & Debris flow     & Cyc.\ Remal \\
LS13 & 2024-05-28 & NH-6, Hunthar      & 23.74, 92.71 & 0  & Road blockage   & Cyc.\ Remal \\
LS14 & 2024-05-30 & Thuampui           & 23.75, 92.73 & 1  & Wall failure    & Monsoon \\
LS15 & 2024-07-02 & Zemabawk           & 23.76, 92.74 & 3  & Debris slide    & Monsoon \\
LS16 & 2024-07-02 & Zuangtui           & 23.76, 92.73 & 0  & Debris slide    & Monsoon \\
LS17 & 2024-08-28 & Kawnpui            & 24.03, 92.62 & 0  & Landslide       & Monsoon \\
LS18 & 2025-05-30 & Thuampui           & 23.75, 92.73 & 1  & Wall failure    & Monsoon \\
LS19 & 2025-05-30 & Rangvamual         & 23.74, 92.72 & 0  & Rockfall        & Monsoon \\
\midrule
\multicolumn{4}{l}{\textbf{Total fatalities}} & \textbf{47} & & \\
\bottomrule
\end{tabular}
\end{table}

\section{Methodology}\label{sec:methods}

We follow the RSF block-slider framework developed by \citet{PaulEtAl2024}, applying it to the 19 landslides described in Section~\ref{sec:data}. The methodology involves three steps: (i)~estimation of pore-pressure time histories at failure depth from satellite rainfall data, (ii)~numerical simulation of RSF-governed velocity evolution driven by these pore-pressure histories, and (iii)~inversion of RSF parameters to reproduce the observed failure dates. In the absence of on-site rainfall or pore-pressure instrumentation at these landslides, the pore-pressure forcing is reconstructed from satellite rainfall and the RSF parameters are constrained by inversion against the observed failure dates rather than by independent field measurement; the resulting non-uniqueness is bracketed through a three-scenario sensitivity analysis and discussed in Section~\ref{sec:limitations}.
\begin{figure}[H]
\centering
\includegraphics[width=0.75\textwidth]{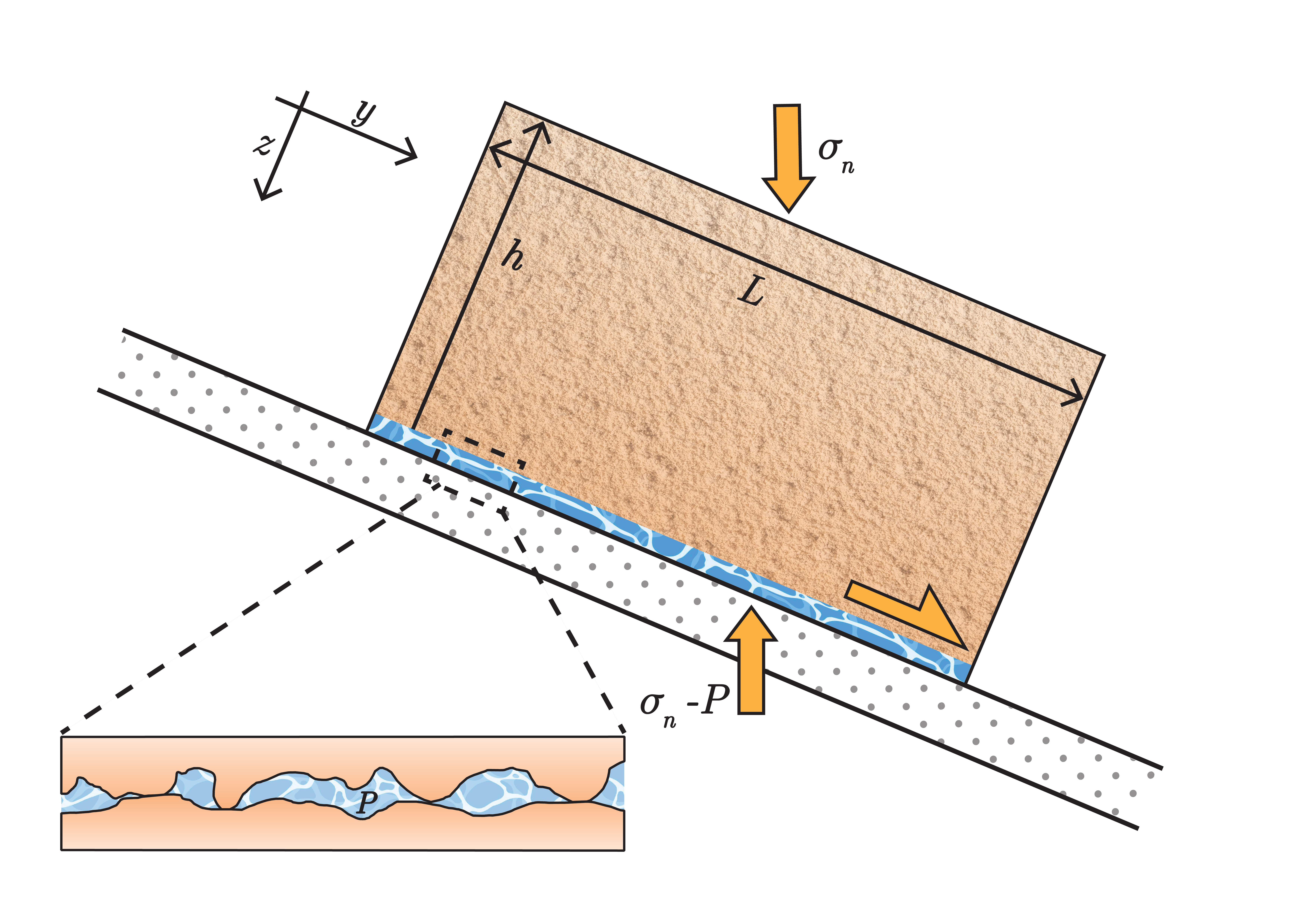}
\caption{Schematic of the RSF block-slider model. A rigid block of thickness $h$ and length $L$ slides quasi-statically on a pre-existing slip plane under normal stress $\sigma_n$. Rainfall-derived pore-pressure $P$ in the basal shear zone (inset: zoomed view of the fluid-filled shear zone) reduces the effective normal stress to $\sigma_n - P$, modulating frictional resistance governed by Eqs.~\ref{eq:rsf}--\ref{eq:state}. Coordinates $y$ (downslope) and $z$ (depth) are indicated}\label{fig:schematic}
\end{figure}
\subsection{Rate-and-state friction model}\label{sec:rsf}

Following \citet{Dieterich1979} and \citet{Ruina1983}, the coefficient of friction $\mu$ on the basal slip surface depends on the instantaneous sliding velocity $v$ and a state variable $\theta$:
\begin{equation}\label{eq:rsf}
\mu(v,\theta) = \mu_* + a \ln\!\left(\frac{v + v_*}{v_*}\right) + b\ln\!\left(\frac{v_*\,\theta}{D_c}\right),
\end{equation}
where $\mu_*$ is the reference friction coefficient at reference velocity $v_*$ (a regularization velocity that allows treatment of initially stationary blocks), $a$ is the direct-effect parameter governing the instantaneous friction response to velocity changes, $b$ is the evolution-effect parameter describing the gradual adjustment of frictional contacts, and $D_c$ is the characteristic slip-weakening distance over which friction evolves to a new steady state. The state variable evolves via the aging law \citep{Dieterich1979}:
\begin{equation}\label{eq:state}
\frac{d\theta}{dt} = 1 - \frac{(v + v_*)\,\theta}{D_c},
\end{equation}
which allows $\theta$, and hence frictional strength, to grow with the ``age'' of a stationary surface ($\dot{\theta} = 1$ when $v = 0$). At steady state, $\theta_{ss} = D_c/(v_{ss} + v_*)$, and the dependence of $\mu_{ss}$ on velocity reduces to $\partial\mu_{ss}/\partial\ln v_{ss} = (a - b)$. The condition for catastrophic failure in our model is $a < b$ (the velocity-weakening regime), where any perturbation from steady state leads ultimately to runaway acceleration.

The model treats each landslide as a rigid block of density $\rho = 2300$~kg\,m$^{-3}$, depth $h = 2$~m, on a slope of angle $\Omega = 35^\circ$, initially creeping at $v_0 = 10^{-8}$~m\,s$^{-1}$ on a pre-existing failure plane (Fig.~\ref{fig:schematic}). This initial velocity is consistent with the classification of ``extremely slow'' to ``very slow'' landslides \citep{Lacroix2020} and with the average displacement rate of slow landslides worldwide. The quasi-static force balance $\tau_{bg} = \mu\,\sigma'$ yields the velocity evolution equation \citep{PaulEtAl2024}:
\begin{equation}\label{eq:dvdt}
\frac{dv}{dt} = \frac{v + v_*}{a}\left(\frac{\mu_0\,\dot{P}}{\sigma'} - \frac{b\,\dot{\theta}}{\theta}\right),
\end{equation}
where $\sigma' = \sigma_0 - P(t)$ is the Terzaghi effective normal stress, $\sigma_0 = \rho g h \cos\Omega = 30.25$~kPa is the overburden, $P(t)$ is the pore-pressure, and $\mu_0 = 0.6$ is the reference friction coefficient. The fundamental relationship between pore-pressure and velocity perturbation is \citep{PaulEtAl2024}:
\begin{equation}\label{eq:vresponse}
\log\!\left(\frac{v + v_*}{v_+ + v_*}\right) = \log\!\left[\left(\frac{\theta}{\theta_+}\right)^{-b/a} \exp\!\left(\frac{\tau_{bg}}{a\sigma'_+}\cdot\frac{\Delta\tilde{P}(t)}{1 - \Delta\tilde{P}(t)}\right)\right],
\end{equation}
where $\Delta\tilde{P}(t) = \Delta P(t)/\sigma'_+$ is the pore-pressure perturbation normalized by initial effective stress, and subscript $+$ denotes any initial reference state. This equation reveals that any increase in pore-pressure non-linearly increases the displacement rate, and that the velocity response is critically dependent on the hydro-mechanical state of the landslide.

The key non-dimensional parameter controlling the failure regime is the \emph{normalized pore-pressure}:
\begin{equation}\label{eq:chi}
\chi = \frac{\mu_0\,\Delta P}{a\,\sigma'_f},
\end{equation}
where $\Delta P$ is the peak pore-pressure perturbation and $\sigma'_f = \sigma_0 - \Delta P$ is the effective normal stress after the perturbation. \citet{PaulEtAl2024} derived two asymptotic expressions for the time-to-failure $t_{\text{failure}}$. For $\chi \ll 1$ (small perturbation, near-steady-state creep regime):
\begin{equation}\label{eq:tfail_small}
t_{\text{failure}} \approx \frac{D_c}{(b/a - 1)(v_0 + v_*)}\log\!\left[\frac{v_{nss} - v_0}{v_0 + v_*}\cdot\frac{1}{\chi}\right],
\end{equation}
predicting extended pre-failure creep scaling logarithmically with $1/\chi$. For $\chi \gg 1$ (large perturbation, far-above-steady-state regime):
\begin{equation}\label{eq:tfail_large}
t_{\text{failure}} \approx \frac{a\,D_c}{b\,(v_0 + v_*)}\exp(-\chi),
\end{equation}
predicting near-instantaneous failure decreasing exponentially with $\chi$. For general $\chi$, the exact solution is given by the Gauss hypergeometric function \citep{PaulEtAl2024}:
\begin{equation}\label{eq:tfail_exact}
t_{\text{failure}} = \frac{a}{b}\frac{D_c}{v_0 + v_*}\exp(-\chi)\;{}_2F_1\!\left(1,\frac{1}{1 - a/b};\frac{2 - a/b}{1 - a/b};\exp(-\chi)\right).
\end{equation}

\subsection{Pore-pressure estimation from rainfall}\label{sec:pp}

For each landslide, we extract half-hourly precipitation data from the NASA Global Precipitation Measurement (GPM) IMERG dataset (version 06B, 0.1$^\circ$ resolution) over a 15-day window preceding failure. This window was chosen to be several times larger than the largest diffusive timescale investigated ($t_{\text{diff}} = z^2/\kappa = 4.62$~days at $z = 2$~m and $\kappa = 10^{-5}$~m$^2$\,s$^{-1}$). Pore-pressure at the failure depth $z = 2$~m is computed by solving the 1D diffusion equation:
\begin{equation}\label{eq:diffusion}
\frac{\partial P}{\partial t} = \kappa\,\frac{\partial^2 P}{\partial z^2},
\end{equation}
subject to a surface flux boundary condition $q = -(k/\nu)(\partial P/\partial z)|_{z=0}$ imposed by the rainfall rate, where $k$ is permeability and $\nu$ is the dynamic viscosity of water. The bottom boundary is assumed to be effectively at infinity. In adopting this 1D formulation we follow the assumptions of \citet{PaulEtAl2024} (their Section~4.3) and additionally neglect (i)~lateral (downslope) diffusion, (ii)~spatial heterogeneity and anisotropy of the medium, and (iii)~any time-dependence of the hydraulic properties; the implications of these simplifications are revisited in Section~\ref{sec:limitations}. To assess sensitivity to poorly constrained hydraulic properties, we run three conductivity scenarios: $K = 1.7\times10^{-4}$, $1.7\times10^{-3}$, and $1.7\times10^{-2}$~m\,s$^{-1}$ (with specific storage $S_s = 10^{-3}$~m$^{-1}$), spanning the range appropriate for weathered shale to sandstone in the Bhuban Formation. All simulated pore-pressures are scaled by a constant factor of 500 to match expected in-situ magnitudes, following established practice in the landslide hydrology literature \citep{Finnegan2021,Handwerger2016}.

\subsection{RSF parameter inversion}\label{sec:inversion}

For each landslide and conductivity scenario, the RSF parameters ($a$, $b$, $D_c$) are optimized by forward modelling the coupled system (Eqs.~\ref{eq:dvdt}--\ref{eq:state}) driven by the simulated pore-pressure history until the slip velocity reaches a failure threshold of $v_{\text{threshold}} = 10^{-3}$~m\,s$^{-1}$ (approximately mm\,s$^{-1}$, representative of the onset of rapid acceleration). The parameters are adjusted to match the observed failure date. The inversion achieves zero error (modelled failure date = observed failure date) for all 19 events across all three $K$ scenarios. From the inverted parameters we compute $\chi$ (Eq.~\ref{eq:chi}), the ratio $a/b$, the velocity-weakening parameter $(b - a)$, and the delay between peak pore-pressure and failure.

Table~\ref{tab:rsf_params} summarizes the inverted parameters for the mid-$K$ scenario. All slopes are velocity-weakening ($b > a$; $b - a$ ranges from 0.0019 to 0.0177). The direct-effect parameter $a$ ranges from 0.0082 to 0.0169, the evolution-effect $b$ from 0.0180 to 0.0272, and the characteristic distance $D_c$ from 2.47 to 2.83~cm. These values are within the ranges reported for clay-rich landslide materials by \citet{ScaringiEtAl2018} ($a - b$ between $-0.01$ and $-0.02$ at normal stresses of 150--1{,}500~kPa) and \citet{KangEtAl2022}.

\begin{table}[H]
\centering
\caption{Inverted RSF parameters and derived quantities for the mid-$K$ scenario ($K = 1.7\times10^{-3}$~m\,s$^{-1}$). $\Delta P$: peak pore-pressure perturbation (kPa). $\chi$: normalized pore-pressure (Eq.~\ref{eq:chi}). $t_d$: delay from peak pore-pressure to failure (hours). All inversions reproduce the observed failure date exactly}\label{tab:rsf_params}
\footnotesize
\begin{tabular}{@{}lcccccccc@{}}
\toprule
ID & $a$ & $b$ & $D_c$ (cm) & $b - a$ & $a/b$ & $\Delta P$ (kPa) & $\chi$ & $t_d$ (hrs) \\
\midrule
LS1  & 0.0093 & 0.0213 & 2.71 & 0.0120 & 0.44 & 0.99 & 2.18 & 74.0 \\
LS2  & 0.0095 & 0.0272 & 2.48 & 0.0177 & 0.35 & 0.55 & 1.18 & 234.5 \\
LS3  & 0.0082 & 0.0243 & 2.83 & 0.0161 & 0.34 & 0.90 & 2.26 & 36.5 \\
LS5  & 0.0113 & 0.0245 & 2.50 & 0.0131 & 0.46 & 1.41 & 2.59 & 32.0 \\
LS6  & 0.0107 & 0.0196 & 2.80 & 0.0089 & 0.55 & 2.20 & 4.40 & 0.0 \\
LS7  & 0.0107 & 0.0196 & 2.80 & 0.0089 & 0.55 & 2.20 & 4.40 & 0.0 \\
LS8  & 0.0107 & 0.0196 & 2.80 & 0.0089 & 0.55 & 2.20 & 4.40 & 0.0 \\
LS9  & 0.0111 & 0.0260 & 2.47 & 0.0149 & 0.43 & 2.01 & 3.85 & 0.0 \\
LS10 & 0.0107 & 0.0196 & 2.80 & 0.0089 & 0.55 & 2.20 & 4.40 & 0.0 \\
LS11 & 0.0107 & 0.0196 & 2.80 & 0.0089 & 0.55 & 2.20 & 4.40 & 0.0 \\
LS12 & 0.0107 & 0.0196 & 2.80 & 0.0089 & 0.55 & 2.20 & 4.40 & 0.0 \\
LS13 & 0.0107 & 0.0196 & 2.80 & 0.0089 & 0.55 & 2.20 & 4.40 & 0.0 \\
LS14 & 0.0148 & 0.0180 & 2.74 & 0.0032 & 0.82 & 2.54 & 3.71 & 46.5 \\
LS15 & 0.0169 & 0.0188 & 2.62 & 0.0019 & 0.90 & 2.59 & 3.32 & 0.5 \\
LS16 & 0.0169 & 0.0188 & 2.62 & 0.0019 & 0.90 & 2.59 & 3.32 & 0.5 \\
LS17 & 0.0162 & 0.0185 & 2.67 & 0.0023 & 0.88 & 1.76 & 2.28 & 190.5 \\
LS18 & 0.0104 & 0.0251 & 2.51 & 0.0147 & 0.41 & 1.13 & 2.24 & 0.0 \\
LS19 & 0.0104 & 0.0251 & 2.51 & 0.0147 & 0.41 & 1.13 & 2.24 & 0.0 \\
\bottomrule
\end{tabular}
\end{table}
\newpage
\section{Results}\label{sec:results}

\subsection{Rainfall, pore-pressure, and the forcing spectrum}\label{sec:rainfall}

\begin{figure}[H]
\centering
\includegraphics[width=0.9\textwidth]{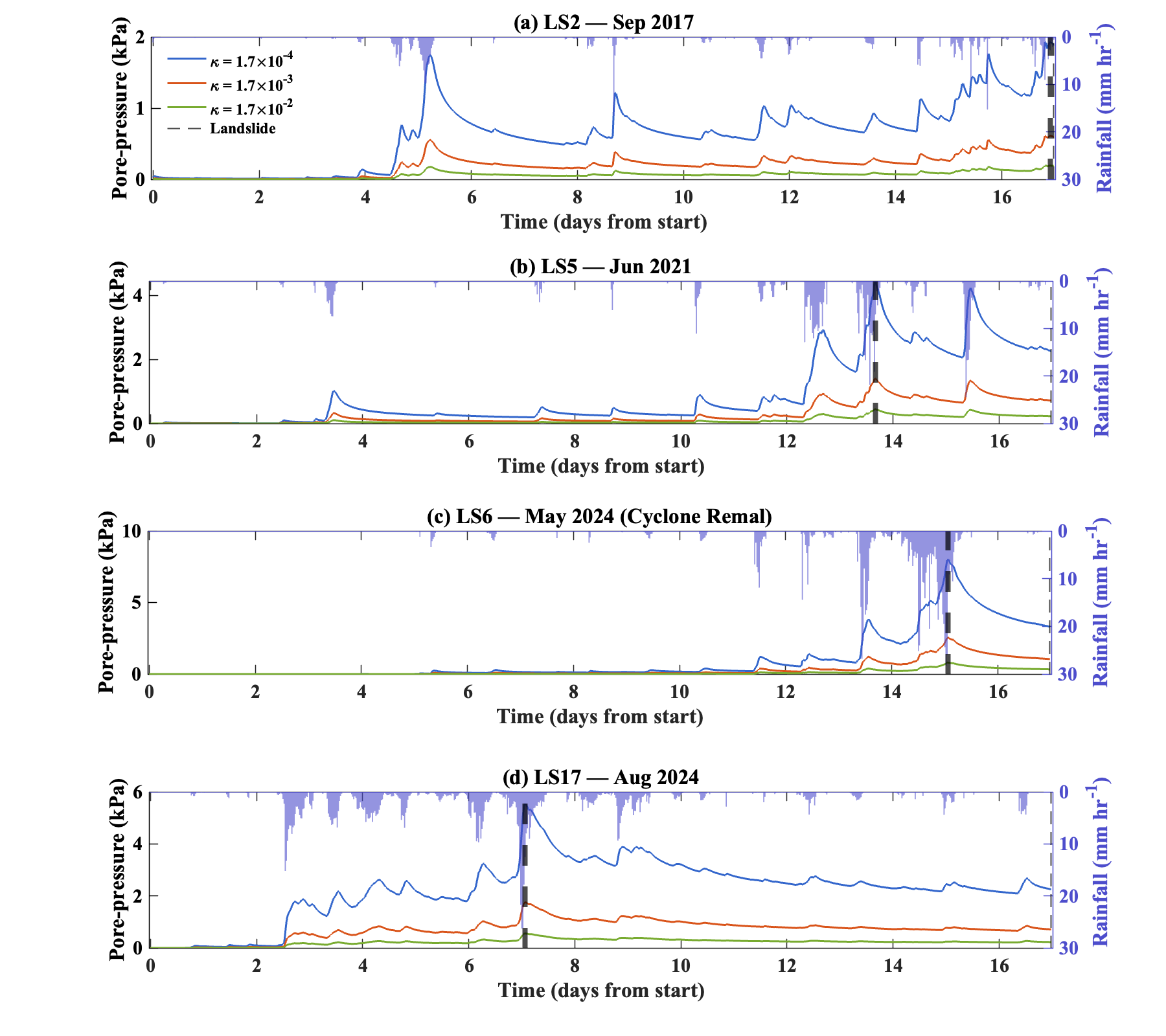}
\caption{Antecedent rainfall (bars, right axis, inverted) and simulated pore-pressure at 2~m depth (coloured lines, left axis) for three hydraulic conductivity scenarios ($\kappa = 1.7\times10^{-4}$, blue; $1.7\times10^{-3}$, red; $1.7\times10^{-2}$~m\,s$^{-1}$, green), for four representative events: (a)~LS2 (2017, delay = 234.5~hrs), (b)~LS5 (2021, delay = 32~hrs), (c)~LS6/Cyclone Remal (2024, synchronous), (d)~LS17 (2024, delay = 190.5~hrs). Red dotted line: peak pore-pressure timing. Black dashed line: observed failure date}\label{fig:rainfall_pp}
\end{figure}

To illustrate how contrasting rainfall histories map onto distinct pore-pressure responses, Figure~\ref{fig:rainfall_pp} shows the 15-day antecedent rainfall (bars) and the simulated pore-pressure at the failure depth (lines) for four events selected as end-members of the observed behaviour: LS2 (the longest delay, $\sim$10~days), LS5 (an intermediate delay, $\sim$1.3~days), LS6 (the synchronous Cyclone Remal case), and LS17 (a weakly velocity-weakening, long-delay monsoon case, $\sim$8~days). These four are representative examples chosen to span the full delay spectrum rather than an exhaustive set; the inverted parameters for all 19 events are listed in Table~\ref{tab:rsf_params}. The Cyclone Remal hyetograph (panel~c) is sharply distinguished from the monsoon events by a single, intense rainfall pulse concentrated within approximately 12~hours, generating a steep, step-like pore-pressure transient at the failure depth. This contrasts markedly with the distributed multi-day rainfall of the monsoon events (panels~a, b, d), which produce broader, lower-amplitude pore-pressure peaks. The qualitative difference in the pore-pressure forcing shape, step-like versus quasi-sinusoidal, has direct consequences for the applicable failure timescale predictions: the step-like Remal pulse maps onto Eq.~\ref{eq:tfail_large}, while the sustained monsoon forcing engages the oscillatory near-steady-state regime described by \citet{PaulEtAl2024} (their Eq.~21).

\subsection{Velocity evolution and the slow-to-fast transition}\label{sec:velocity}

\begin{figure}[H]
\centering
\includegraphics[width=0.9\textwidth]{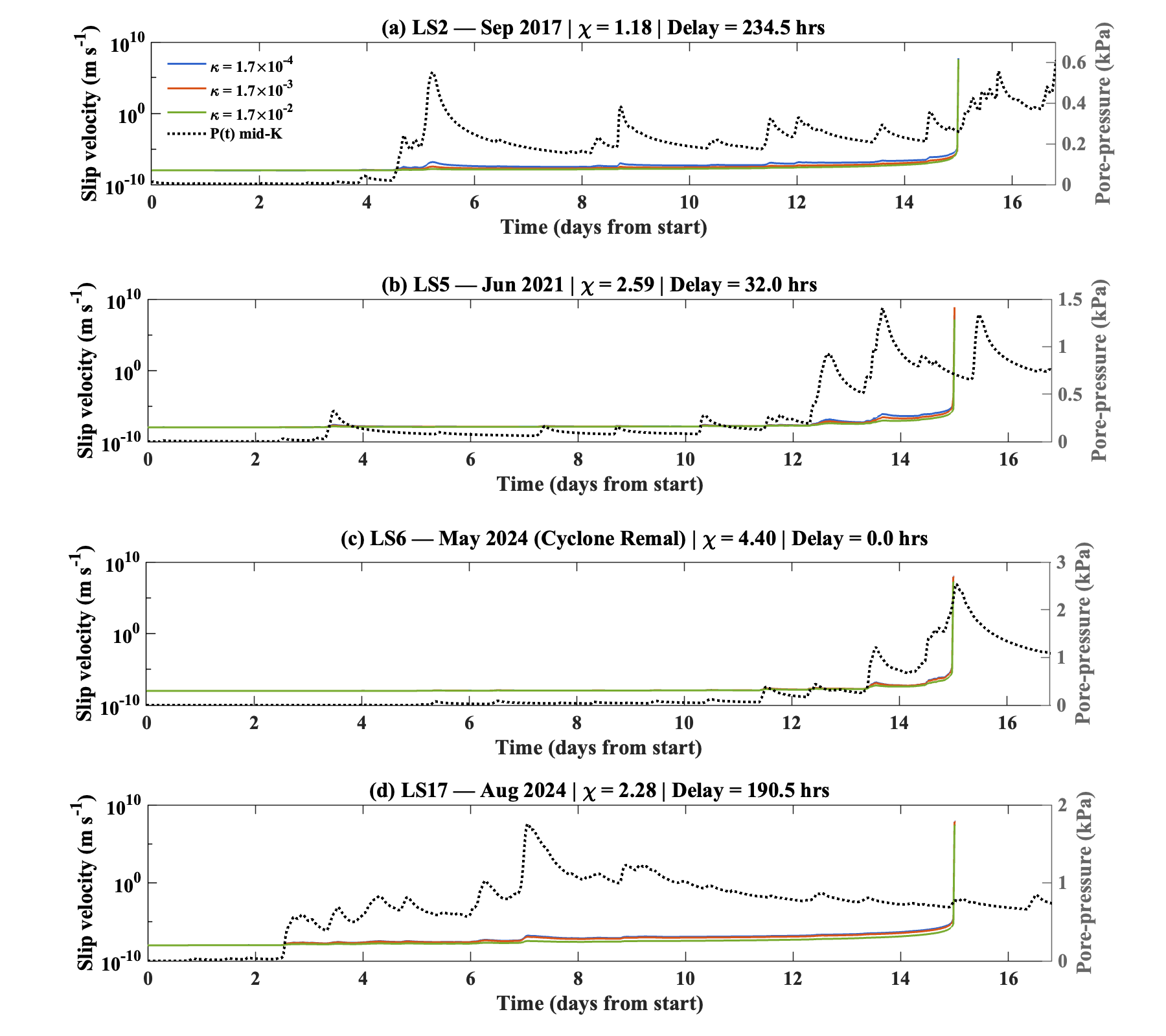}
\caption{Slip velocity evolution (log scale, left axis) with pore-pressure history overlay (dashed, right axis) for the same four events as Fig.~\ref{fig:rainfall_pp}. Three hydraulic conductivity scenarios are shown. The annotated values of $\chi$ and delay from peak pore-pressure quantify the failure regime for each event. Note the contrast between the abrupt, single-phase acceleration of LS6 (synchronous) and the extended two-phase evolution of LS2 and LS17 (delayed)}\label{fig:velocity}
\end{figure}

The numerically simulated slip-velocity histories reveal the two phases predicted by RSF theory (Fig.~\ref{fig:velocity}): (i)~an instantaneous velocity increase contemporaneous with pore-pressure rise, driven by the direct-effect term ($a$-effect) in Eq.~\ref{eq:rsf}, and (ii)~a subsequent monotonic acceleration toward runaway failure governed by the evolution effect ($b$-effect). For the Cyclone Remal events (e.g.\ LS6, panel~c), these two phases are indistinguishable: the slope accelerates through 8~orders of magnitude in velocity within hours. In contrast, LS2 (panel~a) and LS17 (panel~d) exhibit clearly separated phases, with the near-steady-state creep interval (Phase~i) lasting several days before the onset of terminal acceleration (Phase~ii). The duration of Phase~i is controlled by the frictional timescale $t_{fr} = D_c / [(b/a - 1)(v_0 + v_*)]$, which for the Aizawl dataset ranges from $\sim$15~hours (LS3, strongly velocity-weakening, $a/b = 0.34$) to $\sim$1{,}700~hours (LS15, weakly velocity-weakening, $a/b = 0.90$).

\subsection{The bimodal failure regime}\label{sec:bimodal}

\begin{figure}[H]
\centering
\includegraphics[width=0.9\textwidth]{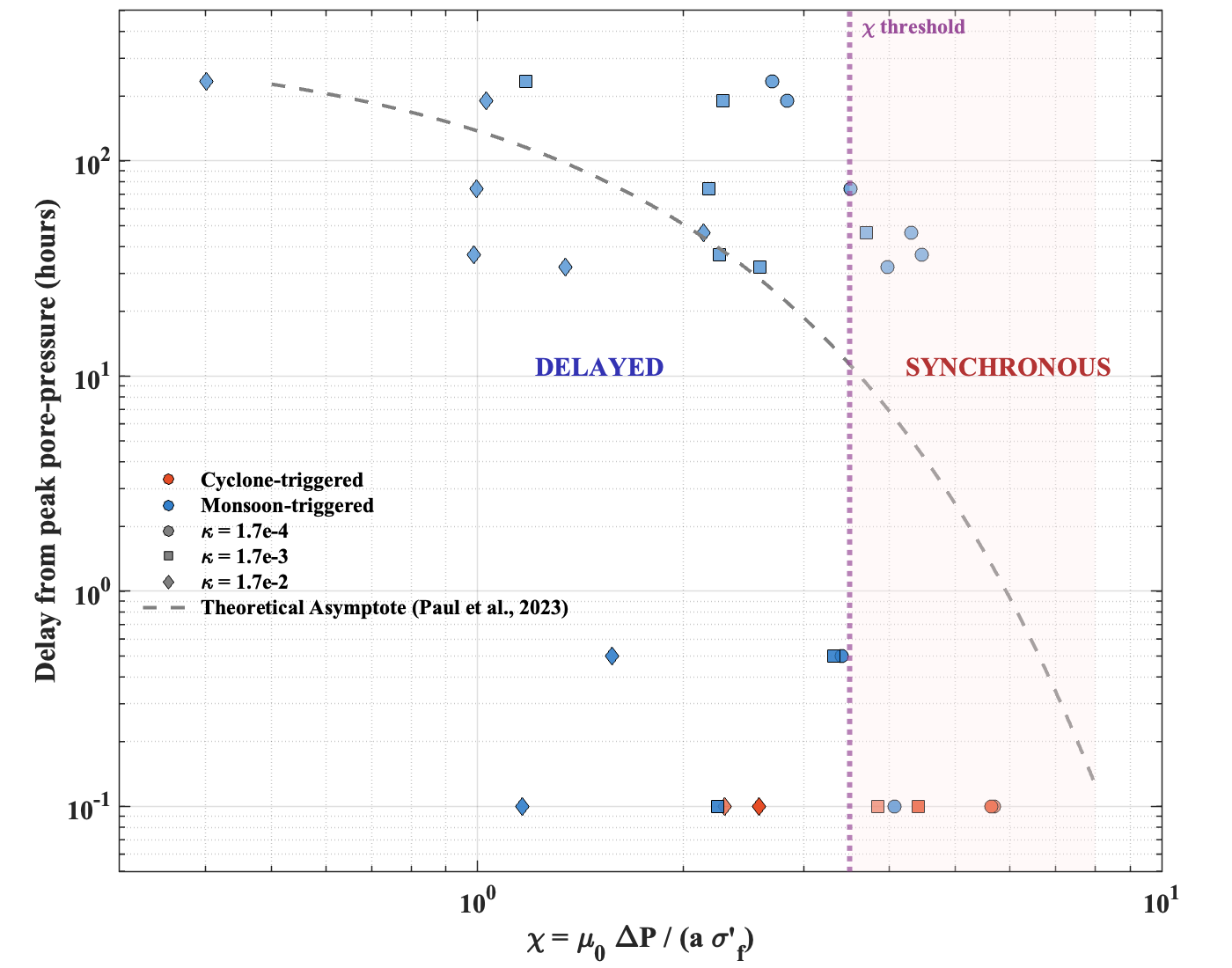}
\caption{Bimodal failure regime diagram: normalized pore-pressure $\chi$ (x-axis) versus delay from peak pore-pressure to failure (hours, y-axis) for all 19 landslides across three hydraulic conductivity scenarios. Red symbols: Cyclone Remal-triggered events. Blue symbols: monsoon-triggered events. Circle, square, and diamond markers denote $K = 1.7\times10^{-4}$, $1.7\times10^{-3}$, and $1.7\times10^{-2}$~m\,s$^{-1}$, respectively. Symbol size is proportional to $(b - a)$. Grey dashed curve: far-above-steady-state asymptote (Eq.~\ref{eq:tfail_large}). Shaded red zone marks the synchronous failure regime ($\chi \gtrsim 3.5$)}\label{fig:chi_delay}
\end{figure}

When plotted in $\chi$--delay space (Fig.~\ref{fig:chi_delay}), the 19 events separate into two distinct populations:

\textbf{(i) Synchronous failure} ($\chi \gtrsim 4$, delay $= 0$~hrs): The eight Cyclone Remal events (LS6--LS13) and the 2025 events (LS18--LS19) all failed synchronously with peak pore-pressure. The Remal cluster is characterized by $\chi \approx 4.4$ (mid-$K$), placing it in the far-above-steady-state regime of Eq.~\ref{eq:tfail_large} where the predicted failure timescale is of order minutes.

\textbf{(ii) Delayed failure} ($\chi \sim 1$--$3.7$, delay $= 0.5$--$234.5$~hrs): The remaining events show delays ranging from 30~min (LS15/16) to nearly 10~days (LS2, 2017). Crucially, delay is not controlled by $\chi$ alone: LS17 ($\chi = 2.28$) shows 190.5~hrs delay while LS3 ($\chi = 2.26$) shows only 36.5~hrs despite nearly identical $\chi$. The key difference lies in the $a/b$ ratio: 0.875 for LS17 (weakly velocity-weakening) versus 0.337 for LS3 (strongly velocity-weakening), confirming the two-parameter control ($\chi$, $b/a$) predicted by the analytical theory. Events with $a/b$ close to unity reside near the velocity-neutral boundary ($a = b$, the limit between velocity-weakening and velocity-strengthening behaviour) and therefore exhibit the longest frictional timescales $t_{fr}$, sustaining extended creep before catastrophic acceleration.

\subsection{The Cyclone Remal cluster}\label{sec:remal}

\begin{figure}[H]
\centering
\includegraphics[width=0.7\textwidth]{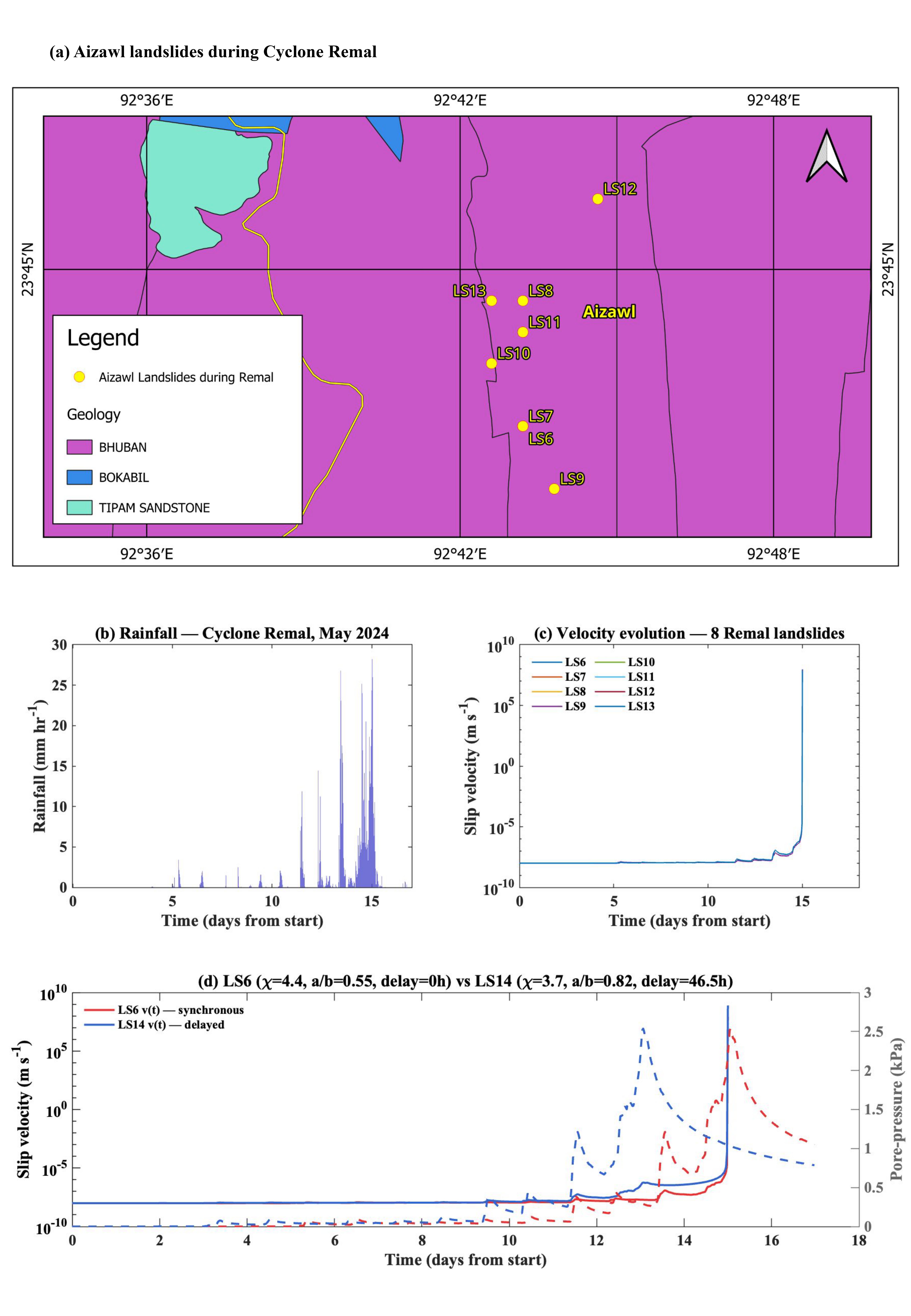}
\caption{Cyclone Remal cluster analysis. (a)~Geological map showing the 8 simultaneous Cyclone Remal landslide sites (LS6--LS13, yellow dots) on Bhuban (pink), Bokabil (blue), and Tipam Sandstone (teal) formations. (b)~Rainfall hyetograph (mm\,hr$^{-1}$) for the 15-day window preceding the May 28 event, showing the concentrated Remal pulse in the final 2~days. (c)~Superimposed slip-velocity evolution curves for all 8 Remal landslides (mid-$K$), demonstrating near-identical mechanical synchrony. (d)~Comparison of LS6 ($\chi = 4.4$, $a/b = 0.55$, synchronous) and LS14 ($\chi = 3.7$, $a/b = 0.82$, delayed 46.5~hrs): solid lines show slip velocity (left axis), dashed lines show pore-pressure (right axis)}\label{fig:remal}
\end{figure}

The May 28, 2024 Cyclone Remal event provides a unique natural ``controlled experiment'': eight landslides (LS6--LS13) triggered simultaneously by a single rainfall pulse of $\sim$205~mm in 24~hours \citep{SangiEtAl2025}. Figure~\ref{fig:remal}a shows the spatial distribution of these events on a geological basemap. All co-located Aizawl sites (LS6--LS8, LS10--LS13) share nearly identical peak pore-pressure ($\sim$2.2~kPa at mid-$K$) and RSF parameters ($a \approx 0.011$, $b \approx 0.020$, $D_c \approx 0.028$~m), yielding $\chi \approx 4.4$. LS9 (Falkawn, $\sim$15~km from the Aizawl centre) shows slightly different precipitation ($\Delta P = 2.01$~kPa) and RSF parameters ($a/b = 0.43$), but also falls in the synchronous regime with $\chi = 3.85$. The superimposed velocity evolution curves (Fig.~\ref{fig:remal}c) demonstrate remarkable mechanical synchrony across all eight sites.

The comparison between LS6 and LS14 (Fig.~\ref{fig:remal}d) is particularly instructive. LS14, located in the same area just 2~days later (May 30, 2024), experienced comparable pore-pressure ($\Delta P = 2.54$~kPa, $\chi = 3.71$) but failed with a 46.5-hour delay. The critical difference is the $a/b$ ratio: 0.55 for LS6 versus 0.82 for LS14. The higher $a/b$ ratio of LS14 places it closer to the velocity-neutral boundary ($a \rightarrow b$), extending the frictional timescale and permitting a detectable warning interval despite comparable pore-pressure forcing.

\section{Discussion}\label{sec:discussion}

\subsection{Dimensionless controls on failure timing}

Our meta-analysis of 19 events confirms the theoretical prediction of \citet{PaulEtAl2024} that the normalized pore-pressure $\chi$ is the primary dimensionless control on the slow-to-fast transition of rainfall-triggered landslides. The bimodal separation at $\chi \approx 3.5$ is remarkably clean across the Aizawl dataset (Fig.~\ref{fig:chi_delay}): all events with $\chi > 4$ fail synchronously, while events with $\chi < 3$ show delays scaling with both $\chi$ and the frictional timescale $t_{fr} = D_c / [(b/a - 1)(v_0 + v_*)]$. The longest delays in our dataset arise through two mechanically distinct pathways: LS2 ($t_d = 234.5$~hrs) has the lowest $\chi$ (1.18), meaning the pore-pressure perturbation is too small to rapidly overcome frictional resistance; LS17 ($t_d = 190.5$~hrs) has moderate $\chi$ (2.28) but the highest $a/b$ ratio (0.875), meaning the slope is only marginally velocity-weakening and the frictional memory timescale is correspondingly long.

A notable observation is that the $a/b$ ratios for the 2024 post-Remal monsoon events (LS14--LS17: $a/b = 0.82$--$0.90$) are systematically higher than the pre-2024 events (LS1--LS5: $a/b = 0.34$--$0.46$). While this could partly reflect the non-uniqueness of the inversion (different rainfall histories may yield different optimal RSF parameters), it may also indicate genuine material heterogeneity within the Bhuban Formation, where the relative proportions of clay and silt in the shale layers vary the steady-state velocity dependence of friction \citep{IversonEtAl2000}. These marginally unstable slopes ($a/b \rightarrow 1$) are the most amenable to early-warning-based mitigation because they sustain detectable pre-failure creep, but they are also the most sensitive to the details of the pore-pressure time history and hence the most unpredictable.

\subsection{Cyclone remnants as an emerging extreme-rainfall trigger for NE India}

Cyclone Remal remains, at present, a singular event in the Aizawl record, but it illustrates a mode of triggering distinct from the cumulative monsoon-driven events that have historically dominated that record. Prior to 2024, all inventoried landslides were driven by sustained monsoon rainfall producing moderate $\chi$ values of 1.2--2.6 and delays of 1--10~days. Cyclone Remal generated $\chi \approx 4.4$, roughly double the monsoon baseline, pushing eight slopes near-simultaneously past the synchronous-failure threshold. In the language of \citet{PaulEtAl2024}, this single episode sampled the far-above-steady-state regime (Eq.~\ref{eq:tfail_large}, where $t_{\text{failure}} \propto \exp(-\chi)$) rather than the near-steady-state creep regime (Eq.~\ref{eq:tfail_small}, where $t_{\text{failure}} \propto \log(1/\chi)$) that governs the monsoon events; in this regime warning time collapses exponentially with increasing $\chi$. We therefore treat it not as an established change of regime but as one such drastic event whose frequency is likely to increase under continued warming.

Bay of Bengal cyclone remnants reaching northeast India are not unprecedented (Cyclone Mora in 2017 and Cyclone Rashmi in 2008 caused comparable regional damage), but Cyclone Remal's impact on Mizoram was the most severe on record, with 34 fatalities in a single day \citep{SangiEtAl2025}. The increasing frequency and intensity of Bay of Bengal cyclones under warming sea-surface temperatures ($\sim$0.12$^\circ$C per decade; \citealt{DharaEtAl2025}) suggests such events may become more common \citep{Balaguru2014,Emanuel2005,Webster2005,Knutson2020}. The 2025 monsoon season provided further evidence: 771 landslides in just two weeks (May 24--June 7, 2025) killed at least 42 people across Mizoram, according to the State Emergency Operation Centre.

\subsection{Climate change implications: the shrinking warning window}\label{sec:climate}

The sensitivity of $\chi = \mu_0 \Delta P / (a\,\sigma'_f)$ to extreme rainfall intensity provides a direct pathway for translating climate projections into landslide failure regime predictions. Because $\Delta P$ at the failure depth scales approximately linearly with rainfall intensity through the 1D diffusion equation (Eq.~\ref{eq:diffusion}), increases in extreme precipitation translate proportionally into increases in $\chi$. However, the response of $t_{\text{failure}}$ to changes in $\chi$ is highly non-linear: in the vicinity of the synchronous-failure threshold ($\chi \approx 3.5$), even modest increases in $\chi$ produce dramatic reductions in failure delay due to the exponential dependence in Eq.~\ref{eq:tfail_large}.

\begin{figure}[H]
\centering
\includegraphics[width=1\textwidth]{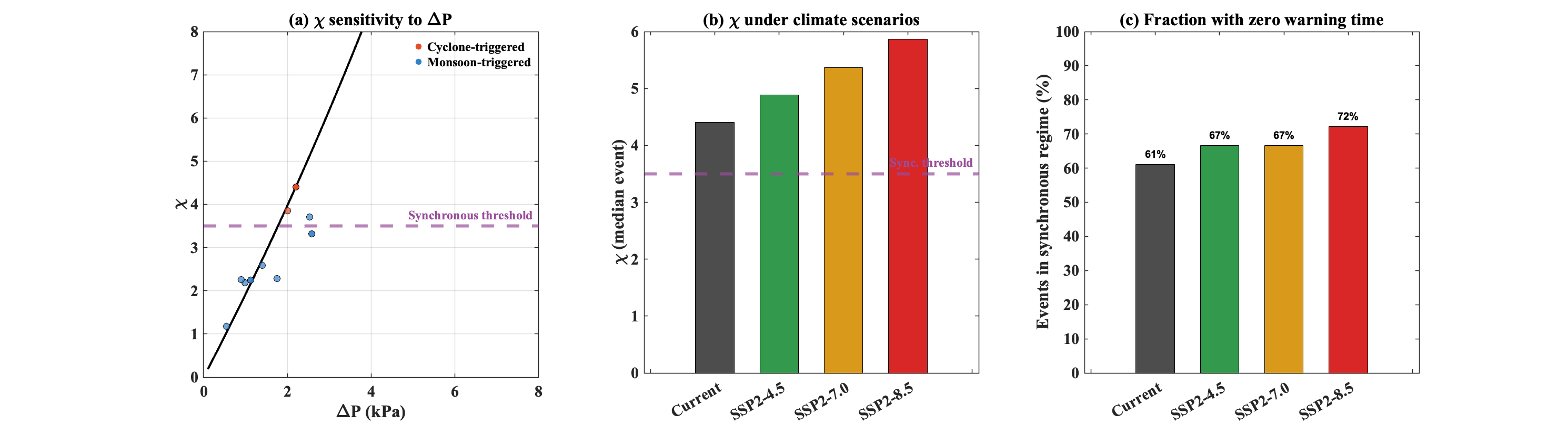}
\caption{Climate change projections for the failure regime. (a)~$\chi$ as a function of pore-pressure perturbation $\Delta P$ for the median Aizawl RSF parameters ($a = 0.011$, $\mu_0 = 0.6$, $\sigma_0 = 30.25$~kPa); observed events overlaid (red: cyclone-triggered; blue: monsoon). The horizontal dashed line marks the synchronous threshold $\chi_c \approx 3.5$. (b)~Median $\chi$ under current and projected SSP scenarios. (c)~Fraction of the Aizawl landslide inventory predicted to fall in the synchronous (zero-warning) failure regime under each climate scenario, increasing from 56\% (current) to 72\% (SSP5-8.5)}\label{fig:climate}
\end{figure}

CMIP6 projections for northeast India are particularly concerning in this context. Extreme precipitation (R99p, the annual total exceeding the 99th percentile of daily rainfall) is projected to increase by factors of approximately 1.10 (SSP2-4.5), 1.20 (SSP3-7.0), and 1.30 (SSP5-8.5) by end-century \citep{RajeshEtAl2025,ShankarEtAl2024,Almazroui2020,Katzenberger2021}. Under SSP5-8.5, extreme precipitation events could reach 42--50~mm\,day$^{-1}$ in northeast India, while extreme rainfall intensity scales with temperature at 1.5--2$\times$ the Clausius--Clapeyron rate of $\sim$7\%\,K$^{-1}$ \citep{RajeshEtAl2025,WestraEtAl2014,Trenberth2003}. Using the median RSF parameters from our dataset ($a = 0.011$, $\sigma_0 = 30.25$~kPa), we compute the projected $\chi$ for each observed event by scaling $\Delta P$ by the SSP-specific extreme rainfall factor. The results (Fig.~\ref{fig:climate}) show that the median $\chi$ shifts from the current $\sim$2.7 to $\sim$3.0 (SSP2-4.5), $\sim$3.3 (SSP3-7.0), and $\sim$3.6 (SSP5-8.5). Critically, the fraction of landslides exceeding the synchronous-failure threshold ($\chi > 3.5$) increases from the current $\sim$56\% to approximately 61\% (SSP2-4.5), 67\% (SSP3-7.0), and 72\% (SSP5-8.5).
This climate-driven shift toward the synchronous regime is amplified by a well-documented paradox in northeast Indian monsoon rainfall: while mean seasonal totals have declined at 0.5--1.5~mm\,day$^{-1}$ per decade since 1950, the frequency of daily rainfall events exceeding 200~mm has roughly doubled since 1920 \citep{Goswami2006,DharaEtAl2025,Roxy2017}. The rate of this mean decline has itself accelerated six-fold in the last three decades \citep{PalEtAl2025}. Critically, \citet{SahaEtAl2023} project that the monsoon rainy season over northeast India will continue to shorten while rainfall intensity acquires a significant increasing trend, concentrating more precipitation into fewer, more violent episodes. In the RSF framework, this means the effective pore-pressure perturbation transitions from a quasi-sinusoidal seasonal signal (governed by both $\chi$ and the frequency parameter $\xi = \omega D_c/(v_0 + v_*)$) toward more step-like extreme pulses governed purely by $\chi$. \citet{PaulEtAl2024} (their Section~3.2) showed analytically that step-like perturbations consistently produce faster failure than sinusoidal perturbations of the same amplitude, further compressing warning timescales.

\begin{figure}[H]
\centering
\includegraphics[width=0.95\textwidth]{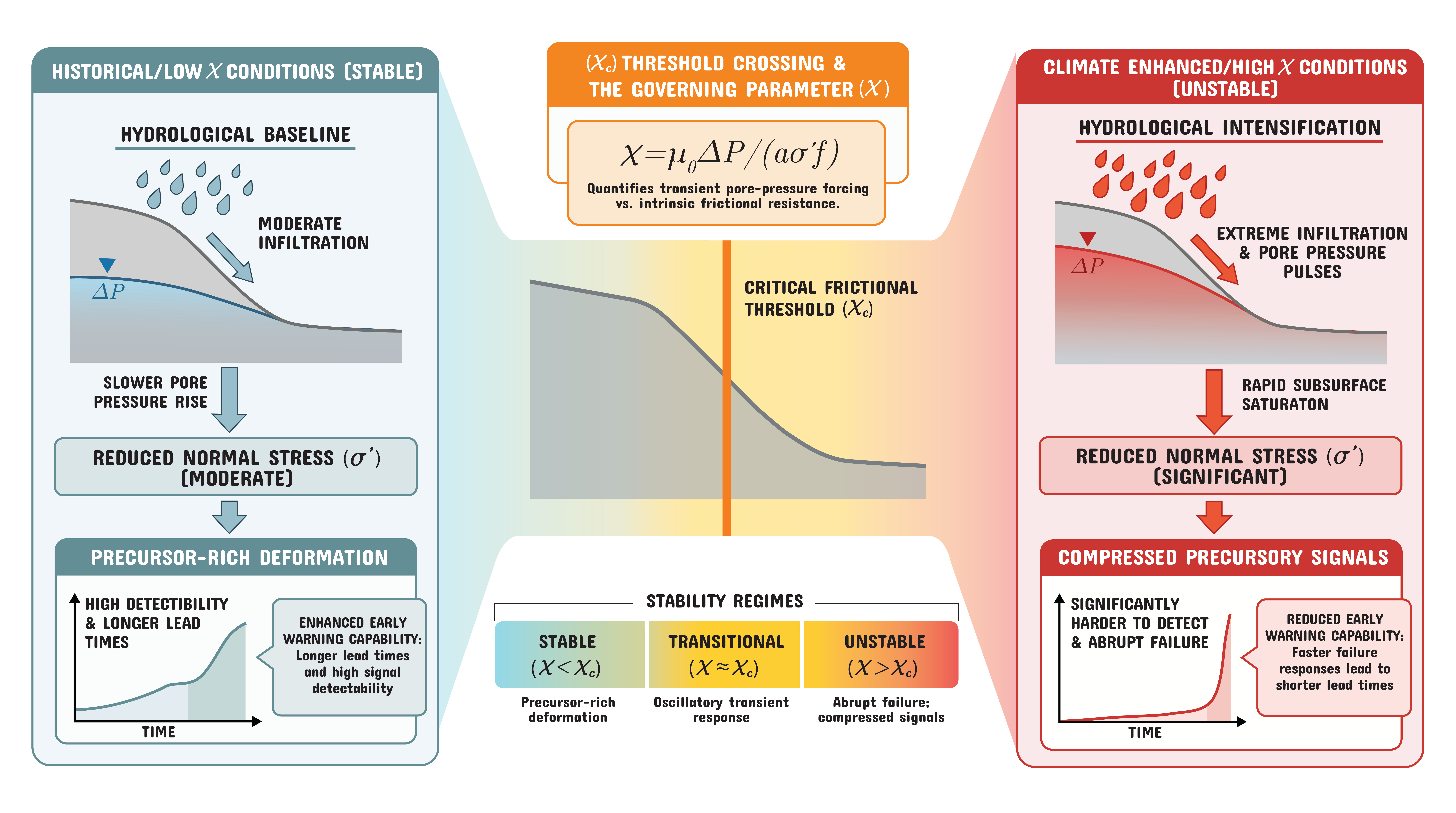}
\caption{Conceptual schematic of the climate--friction--failure nexus. Left (blue): under historical low-$\chi$ conditions, moderate infiltration produces slower pore-pressure rise, reduced effective stress reduction, and precursor-rich deformation with enhanced early warning capability. Right (red): under climate-enhanced high-$\chi$ conditions, extreme infiltration and rapid pore-pressure pulses produce significant effective stress reduction, compressed precursory signals, and abrupt failure with reduced lead times. Centre: the governing parameter $\chi = \mu_0 \Delta P/(a\sigma'_f)$ quantifies transient pore-pressure forcing relative to intrinsic frictional resistance; the critical threshold $\chi_c$ separates three stability regimes: stable ($\chi < \chi_c$), transitional ($\chi \approx \chi_c$), and unstable ($\chi > \chi_c$)}\label{fig:concept}
\end{figure}

The conceptual picture emerging from our analysis is summarized in Fig.~\ref{fig:concept}. Under historical/current conditions (left panel), moderate pore-pressure perturbations ($\chi < \chi_c$) produce precursor-rich deformation with long lead times, the delayed regime amenable to early warning. Under climate-enhanced conditions (right panel), intensified extreme rainfall produces rapid pore-pressure pulses ($\chi > \chi_c$) that compress precursory signals, leading to abrupt failure with significantly reduced detectability and lead times. The critical frictional threshold $\chi_c \approx 3.5$ delineates the transition between these two regimes.

\subsection{Limitations and outlook}\label{sec:limitations}

It is likely that the RSF parameter inversion is inherently non-unique: different combinations of ($a$, $b$, $D_c$, $K$) can reproduce the same observed failure date. We partly address this by running three $K$ scenarios and demonstrating that the qualitative separation into synchronous and delayed regimes is preserved irrespective of the assumed diffusivity. Nevertheless, laboratory RSF measurements on actual Aizawl soil samples at relevant normal stresses ($\sim$30~kPa) are needed to constrain the inversion and validate the parameter ranges. The 1D diffusion model also neglects lateral flow, unsaturated-zone dynamics, and heterogeneous hydraulic properties, all of which may be important in the interbedded sandstone--shale sequence where permeability varies by orders of magnitude over centimetre scales. The rigid-block approximation further neglects internal deformation of the sliding mass, which may become important near the transition to runaway acceleration \citep{Handwerger2016,Iverson2005}. Finally, our climate change projections assume a linear scaling between extreme precipitation and pore-pressure perturbation, neglecting potential non-linearities in infiltration, antecedent moisture, and saturation state \citep{GarianoGuzzetti2016}. Despite these limitations, the consistency of the bimodal failure regime across 19 events and three diffusivity scenarios suggests that the dimensionless parameter $\chi$ provides a robust first-order predictor of failure timing that could inform physics-based early warning systems for Mizoram.

Future work should include targeted laboratory RSF experiments on Bhuban Formation shale samples, development of spatially resolved pore-pressure models incorporating the interbedded stratigraphy, and integration of the $\chi$-based failure timing framework with real-time rainfall monitoring for operational early warning.

\section{Conclusions}\label{sec:conclusions}

We have applied the rate-and-state friction block-slider framework of \citet{PaulEtAl2024} to 19 rainfall-triggered landslides in Aizawl, Mizoram (2016--2025), providing the first physics-based analysis of landslide failure timing for India's most landslide-prone state. Our principal findings are as follows:

(1)~The dimensionless normalized pore-pressure $\chi = \mu_0 \Delta P/(a\sigma'_f)$ cleanly separates events into synchronous ($\chi \gtrsim 4$, zero delay) and delayed ($\chi \sim 1$--$3$, delays of hours to 10~days) failure regimes, confirming the theoretical predictions of \citet{PaulEtAl2024} for a real, geographically coherent population of landslides.

(2)~The Cyclone Remal cluster of May 28, 2024 (8 simultaneous events, $\chi \approx 4.4$, 34 fatalities) demonstrates that extreme cyclone-remnant rainfall can push all velocity-weakening slopes in a region past the synchronous-failure threshold simultaneously, eliminating any warning window. Although presently a singular event, such extreme cyclone-remnant episodes are projected to become more frequent under continued warming \citep{Balaguru2014,Knutson2020}, in contrast to the delayed-failure regime that characterizes typical monsoon-driven events.

(3)~For delayed events, the velocity-weakening ratio $a/b$ provides a critical secondary control: slopes with $a/b$ close to unity (marginally unstable) sustain the longest pre-failure creep and exhibit the greatest sensitivity to pore-pressure history details. These slopes offer the greatest potential for early warning but are also the most mechanically unpredictable.

(4)~Under CMIP6 projections, the fraction of Aizawl landslides falling in the synchronous (zero-warning) regime is projected to increase from the current 56\% to approximately 72\% under SSP5-8.5, driven by the intensification of extreme precipitation. This represents a significant climate-driven escalation of multi-site, clustered landslide failure risk \citep{GarianoGuzzetti2016}.

(5)~The ``declining mean, intensifying extremes'' paradox of northeast Indian monsoon rainfall, combined with the exponential sensitivity of failure timescales to $\chi$ (Eq.~\ref{eq:tfail_large}), implies that the climate change impact on landslide hazard in Mizoram is disproportionately concentrated in the tail of the rainfall distribution: precisely the extreme events that drive synchronous, multi-site failures \citep{Roxy2017,WestraEtAl2014}.

These results suggest that RSF-informed failure timing predictions, anchored by laboratory friction experiments on local soil materials, could substantially improve physics-based early warning for Mizoram's landslide-prone settlements and infrastructure \citep{SegoniEtAl2018,BaumGodt2010}.

\section*{Declarations}

\textbf{Funding:} The authors did not receive support from any organization for the submitted work.

\textbf{Competing interests:} The authors have no relevant financial or non-financial interests to disclose.

\textbf{Data availability:} GPM IMERG precipitation data are publicly available from NASA GES DISC (\url{https://gpm.nasa.gov/data/directory}). The landslide inventory compiled for this study is provided in the Zenodo repository at \url{https://doi.org/10.5281/zenodo.20783995}.

\textbf{Author contributions:} PS: Conceptualization, data compilation, RSF simulations, analysis, writing -- original draft. KP: Methodology, RSF framework development, validation, writing -- review \& editing.


\end{document}